\documentclass[%
superscriptaddress,
 amsmath,amssymb,
 aps,
 prl,
 reprint,
floatfix,
]{revtex4-2}
\usepackage{graphicx,xcolor}
\usepackage{ulem}
\usepackage{pifont}
\definecolor{darkblue}{RGB}{0,0,150}
\definecolor{nightblue}{RGB}{0,0,100}

\usepackage{mathrsfs,dsfont,mathtools}

\usepackage{dcolumn}% Align table columns on decimal point
\usepackage{bm}% bold math
\usepackage[
colorlinks,
citecolor=darkblue,
linkcolor=darkblue,
urlcolor=nightblue]{hyperref}% add hypertext capabilities
\usepackage[english]{babel}
\usepackage[babel,kerning=true,spacing=true]{microtype}

\usepackage{feynmp-auto}
\usepackage{MnSymbol}

\newcommand{\bff}[1]{\mathbf{#1}}

\definecolor{DarkRed}{RGB}{100,0,0}
\definecolor{DarkBlue}{RGB}{000,0,100}

\newcommand{\xmark}{\text{\ding{55}}}
\AtBeginDocument{\renewcommand{\natexlab}[1]{#1}}% <--- the fix

\newcommand{\pv}[1]{{\color{black}#1}}

\begin{document}

\title{Quantum geometric photocurrents of quasiparticles in superconductors}

\author{Daniel Kaplan}
\email{d.kaplan1@rutgers.edu}
\affiliation{%
 Department of Physics and Astronomy, Center for Materials Theory,
Rutgers University, Piscataway, NJ 08854, USA
}

\author{Kevin P.~Lucht}
\affiliation{%
 Department of Physics and Astronomy, Center for Materials Theory,
Rutgers University, Piscataway, NJ 08854, USA
}%
\author{Pavel A.~Volkov}%
\affiliation{%
Department of Physics, University of Connecticut, Storrs, Connecticut 06269, USA
}%
\author{J.~H.~Pixley}
\affiliation{%
 Department of Physics and Astronomy, Center for Materials Theory,
Rutgers University, Piscataway, NJ 08854, USA
}%
\affiliation{
Center for Computational Quantum Physics, Flatiron Institute, 162 5th Avenue, New York, NY 10010
}%

%\date{}
\begin{abstract}
Nonlinear optical response is a sensitive probe of the geometry and symmetry of electronic Bloch states in solids. Here, we extend this notion to the 
Bogoliubiov-de-Gennes (BdG) quasiparticles in superconductors. We present a theory of photocurrents in superconductors and show that they sensitively depend on the quantum geometry of the BdG excitation spectrum.
For all light polarizations, the photocurrent is proportional to the quantum geometric tensor: for linear polarized light it is related to the quantum metric and for circular polarization --- the Berry curvature dipole of the associated BdG bands.We further relate the photocurrent to the ground state symmetries, providing a symmetry dictionary for the allowed photocurrent responses.
For light not at normal incidence to the sample, photocurrent probes time-reversal symmetry breaking in systems with chiral point groups (such as twisted bilayers). We demonstrate that photocurrents allow to probe topology and TRS breaking in twisted $d$-wave superconductors and test the nature of superconductivity in twisted WSe$_2$ and multilayer stacks of rhombohedral graphene.
Our results pave the way to contactless measurement of the quantum geometric properties and symmetry of superconductivity in materials and heterostructures.

\end{abstract}

\maketitle

The rapidly expanding field of nonlinear optical response has opened the door to measure properties of materials that have been hitherto out of reach. Importantly, the modern description of topological band theory has found connections between the nonlinear response and the quantum geometry of the electronic wavefunction creating intense theoretical and experimental interest~\cite{Ma2021_review, Ma2023_review}. Photocurrents permit spatially sensitive, polarization and frequency tunable charge responses in both insulators and metals \cite{Reserbat_Plantey_2021}. Their measurement plays an important role in the characterization of quantum materials, such as in three-dimensional (3D) Weyl semimetals \cite{Wu2017,Osterhoudt2019}, 2D twisted bilayer graphene (TBG)~\cite{Ottender2020,Hesp2021,Hubmann2022,Kaplan2022,Chaudhary2022shift,kumar2024terahertzphotocurrentprobequantum} and transition metal dichalcogenides (TMDs)~\cite{Xie2016}. Of particular significance are photogalvanic responses, which generate a dc current (or voltage) linear in the intensity of light incident at frequency $\omega$ on a material. 
The resonant dc photocurrent has been previously connected with the Berry curvature of topological materials producing a \pv{quantized photogalvanic response}
\cite{dejuan2017quantized}. In low dimensional systems with time-reversal symmetry (TRS), the nonlinear Hall response is intimately tied to band topology, and is proportional to the Berry curvature dipole on the Fermi surface \cite{kang2019nonlinear,ma2019observation}, while in TRS-broken systems, the dipole of the quantum metric creates both  longitudinal and Hall nonlinear responses, which are independent of scattering time \cite{Gao2014field,gao2023quantum,wang2023quantum,kaplan2024unification}. Thus, it is clear that for charge currents, nonlinear transport unlocks the topological and geometric properties of Bloch electrons \cite{Ahn_2021} that are particularly sensitive to crystal and global symmetries, such as TRS. 

Extensions of these ideas to measure the \pv{quantum geometric properties of charge-neutral \cite{kivelson_1990}  quasiparticles in superconductors, is of the utmost importance, since usual charge transport probes of topology (e.g., Hall effect) are inapplicable in superconductors}. Identifying quantum materials that realize the topological superconductivity (required for fault-tolerant quantum computation \cite{preskill1997faulttolerantquantumcomputation, Kitaev_2003}) requires an accurate determination of the symmetries and ground state of the superconductor \cite{Nayak2008review}. Thus, non-invasive and sensitive probes of the quantum state are imperative \cite{Mong2014Universal}.  
In particular, the usual prescription for realizing $\mathbb{Z}$-topological superconductivity requires the  breaking of TRS in the superconducting state  \cite{Sau2011,stern2013topological}, but even the detection of this symmetry breaking remains challenging in general~\cite{zhao2023time,Volkov_2024}. The ``smoking-gun'' technique to determine the topological states involves measuring the thermal Hall effect  \cite{Kane1997,Vishwanath2001,Hu_2023} -- a typically complicated and difficult experiment, especially for 2D materials. 
With the long-standing search for topological superconductors, new experimental probes that detect the nature of the superconducting wavefunctions' quantum geometry while unambiguously determining their underlying symmetries through optical means, would represent a fundamentally new approach to this long-standing problem.

\begin{figure}[t]
    \centering
\includegraphics[width=.95\columnwidth]{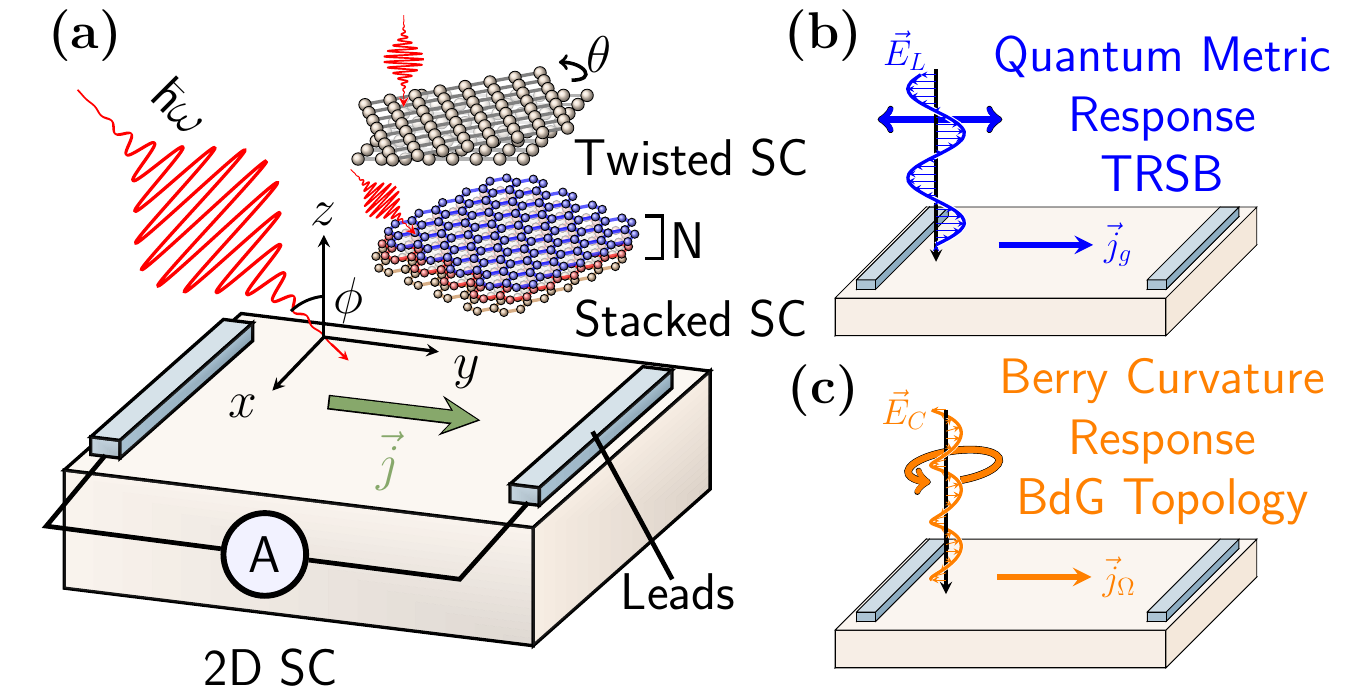}
    \caption{
    {\bf The injection photocurrent in topological superconductors.} 
    (a) Light, at a THz frequency $\omega$, is shown by the red pulse applied either at off-normal incidence to the $x-y$ plane by an angle $\phi$ or normal $(\phi = 0)$ onto a 2D superconductor. In response to the incident light, the superconductor can generate a current shown by $\vec{j}$ represented by the green arrow. This 2D superconductor is general, and could be for example a twisted bilayer with light applied normal to the surface,  or a $N$ layer rhombohedral graphene stack,  pictured above. (b) and (c) show  the main mechanisms of the photocurrent response for different light polarizations.
    %BdG quasiparticles interact with light via its polarization. 
    In (b) [(c)], linearly [circularly] polarized light interacts via the quantum geometric tensor $Q^{\alpha \beta}$ [Berry curvature tensor $\Omega^{\alpha \beta}$] producing a photocurrent $\vec{j}_\textrm{g}$ [$\vec{j}_\Omega$] in the presence of TRSB. The typical scale for the resonant response involves the interlayer tunneling in 2D stacks, $t$.
    }
    \label{fig:fig1}
\end{figure}

In this work, we propose probing the quantum geometry of %a topological 
superconductors through photocurents that appear as a nonlinear response to an external electromagnetic field. 
The latter produces resonant excitations of the BdG bands, creating a normal (non-superconducting) current, see Fig.~\ref{fig:fig1}(a) for a schematic set up. In the following, we present a formalism that evaluates this DC photocurrent from an effective (i.e. BdG) description of a superconductor. As an essential finding of general use, we show how the photocurrent is directly related to two fundamental aspects of quantum geometry of the BdG bands: the quantum metric and the Berry curvature -- accessible separately by different incident light polarization, see Fig.~\ref{fig:fig1}(b) and (c). In addition to the in-plane photocurrent that is produced in a two-dimensional superconducting system, out-of-plane polarization is generated making this a useful tool in the engineering of displacement fields in a gate-free fashion. %In particular, 
Finally, we show that photocurrents serve as a sensitive probe of the symmetry of the superconducting state. As one example, we predict the signatures of broken TRS in photocurrent response.

%As an example,} we isolate currents that directly probe broken TRS in the superconducting state. 

We apply this formalism to propose an experimental probe of the quantum geometry and symmetries of the BdG band structure of twisted bilayers of nodal %$d$-wave 
superconductors (TBSCs), applicable to the  van der Waals high-temperature superconducting material Bi\textsubscript{2}Sr\textsubscript{2}CaCu\textsubscript{2}O\textsubscript{8+$\delta$} (Bi-2212)
\cite{zhao2023time}.
Using this insight we propose probing the %full 
phase diagram and TRS breaking in %the 
twisted Bi-2212 heterostructures, as well as offer a prescription to resolve the chiral nature of superconductivity in twisted WSe\textsubscript{2} (tWSe$_2$) \cite{guerci2024topologicalsuperconductivityrepulsiveinteractions} and multilayer rhombohedral graphene devices \cite{yang2024diverseimpactsspinorbitcoupling,han2024signatureschiralsuperconductivityrhombohedral}.

\paragraph{Theory of quasiparticle photocurrents in superconductors ---}
Our starting point is a theory of quasiparticles in superconductors, described by an effective Hamiltonian.
We follow and simplify the framework previously presented in Refs. \cite{Xu_2019,Papaj_2022,Watanbe2022nonreciprocal,Tanaka2023,Raj_2024}. The effective Hamiltonian is given in the BdG form \cite{de2018superconductivity,Bogoljubov1958},  
\begin{align}
    H =\sum_{\mathbf{k}}\Psi^\dagger_\mathbf{k} \hat{h}_{\textrm{BdG}} \Psi_\mathbf{k}, \,\,\, h_{\textrm{BdG}}(\mathbf{k}) = \left(\begin{matrix}
H_0(\mathbf{k}) & \Delta_\mathbf{k} \\ 
\Delta_\mathbf{k}^\dagger & -H_0(-\mathbf{k})
\end{matrix}\right)
\end{align}
$H_0(\mathbf{k})$ is the normal state Hamiltonian and $\Psi_\mathbf{k}$ is the Gorkov-Nambu spinor.
Under minimal coupling, the Hamiltonian is transformed between particle and hole sectors as $h_\textrm{BdG} \to h_{\textrm{BdG},\mathbf{A}} = \left(\begin{smallmatrix}
H_0(\mathbf{k}+e\mathbf{A}) & \Delta_\mathbf{k} \\ 
\Delta_\mathbf{k}^\dagger & -H_0(-(\mathbf{k}-e\mathbf{A}))
\end{smallmatrix}\right)$. One can then systematically derive the current operator,
\begin{align}
    \mathbf{J}^a_{\textrm{BdG}}(\mathbf{k}) = -\frac{\delta h_{\textrm{BdG}}}{\delta A^a} = -e\hat{v}^a - \frac{e^2}{2}\sum_b\frac{\partial^2 \pv{h}_{\textrm{BdG}}}{\partial A_a \partial A_b} A_b+ \ldots,
\end{align}
where  $\hat{v}^a = \left. \frac{\pv{\partial h_{\textrm{BdG}}}(k-e\mathbf{A} \tau_3/\hbar)}{\partial A_a}\right|_{A\to 0,\Delta_\mathbf{k} \to 0}$ is the effective velocity.
The macroscopic current is then calculated via standard perturbative techniques (see SM):
\begin{align}
    j^a = \frac{1}{i (2\pi)^{3}} \int \textrm{d}E \textrm{d}^2k\textrm{Tr}(G(E,\mathbf{k})\mathbf{J}_{\textrm{BdG}}^a).  
\end{align}
Here and in the following we focus on $d=2$ dimensions.
Furthermore, we will not take into account the corrections to $\hat{v}^a$ from the gap function \cite{Oh2024}, that arise from pair hopping interactions. The Green's function $G(E,{\bf k})$ (in frequency $E$ and momentum ${\bf k}$) is expanded  to second order (details in SM) in ${\bf A}$, which is related to the physical field $\mathbf{E}$ via $\mathbf{E} = - \partial_t \mathbf{A}(t)$.
We model the electromagnetic field with polarization ${\bf E} = (E_x, E_y, E_z)$ and we allow for both linear and circular polarization. 
The electric field has the following frequency structure: $\mathbf{E}(\omega) = \sum_{p=\pm} \mathbf{E}_0 \left(\delta(\omega+p\omega_0)\right)$. At second order, we expect divergences which we regularize by shifting one delta function with a small $\Omega$ (see SM). Focusing on the leading term in the limit $\Omega \to 0$, $\omega_0$ is the central frequency of the light pulse. Thus, the current response we consider is of the so-called injection type \cite{holder2020consequences,kaplan2024unification,Watanbe2022nonreciprocal} (with other contributions discussed below and in the SM). The injection contributions corresponds to a constant time derivative of the current.

Focusing on low temperatures, we find that the leading order contribution results in a ballistic injection of quasiparticle current, $\frac{\partial j^a}{\partial t}$, and is given by,
\begin{align}
     \frac{\partial j^a}{\partial t} = -\frac{ e^3 E^\alpha_\omega E^\beta_{-\omega}}{2\pi\hbar} \int \mathrm{d}^2 k \delta(\hbar\omega_0 - \Delta_{he})\mathcal{D}^a_{he}Q^{\alpha \beta}_{h e},
     \label{eq:jqtensor}
\end{align}
where $a,\alpha,\beta$ denote cartesian directions for the current and electric field polarization, respectively. $Q^{\alpha \beta}_{he}$ is the quantum geometric tensor \cite{Ahn_2021} of the BdG bands defined as $Q^{\alpha \beta}_{he} = \frac{1}{2}\mathcal{A}^a_{he}\mathcal{A}^b_{eh} = g^{ab}_{he}-\frac{i}{2}\Omega^{ab}_{he}$. Here $g^{ab}$ is the quantum metric for BdG bands, $g^{ab}_{he} = \frac{1}{2}\mathcal{A}^a_{he}\mathcal{A}^{b}_{eh} + \textrm{c.c.}$ and their Berry curvature is $ \Omega^{ab}_{he} = i(\mathcal{A}^a_{he}\mathcal{A}^{b}_{eh} - \textrm{c.c.})$. The Berry connection for the particle-hole pairs is $ \mathcal{A}^a_{he} = \frac{\hbar}{e}\left. \langle h (\mathbf{k}-e\mathbf{A}/\hbar \tau_3) | \frac{\partial}{\partial{A_a}} e(\mathbf{k}-e\mathbf{A}/\hbar \tau_3)\rangle\right|_{A \to 0,\Delta_k \to 0}$. 
$e, h$ refer to the (energy $>0$), (energy $<0$) bands of the BdG Hamiltonian. All states below zero energy are assumed occupied at low temperatures, while all quasiparticles states with energies above $\mu$ are empty. 

\begin{figure}
    \centering
\includegraphics[width=1.0\columnwidth]{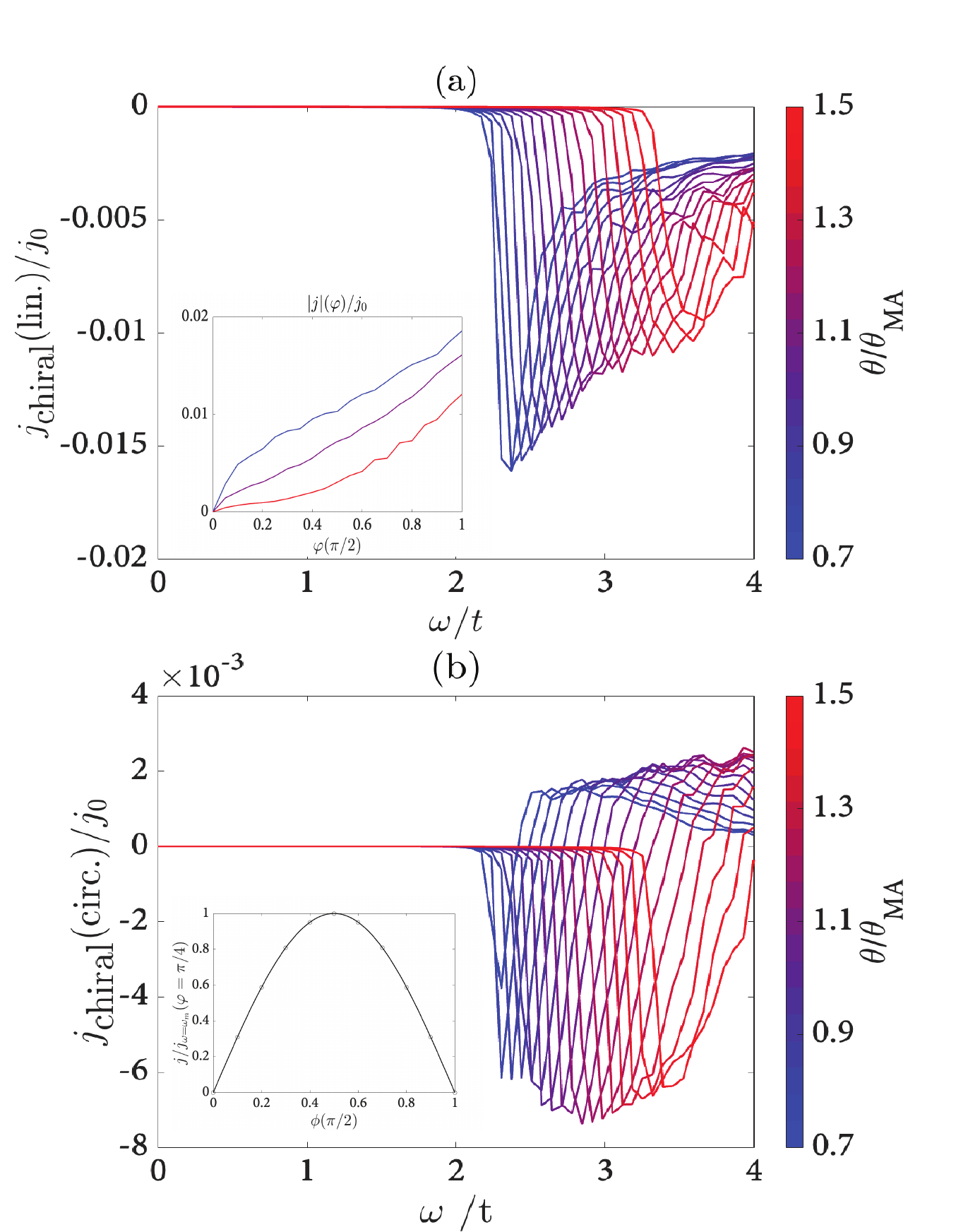}
    \caption{{\bf Off-normal response of the chiral current for TBSCs}:  We consider electric fields tilted at an angle $\phi$  with respect to the sample normal [depicted in Fig.~\ref{fig:fig1}(a)], for the  Hamiltonian in Eq.~\eqref{eq:Ham}, $\delta H_{\textrm{sym}} = 0$. (a) Quantum metric current $g^{\alpha \beta}$ in response to linear polarized light. (b) Berry curvature $\Omega^{\alpha \beta}$ current in response to circularly polarized light. Both are plotted as a function of twist angle (color scale). Inset of (a) shows the magnitude of the photocurrent for three twist angles $\theta/\theta_{MA} = 0.7, 1.0, 1.5$ (blue, purple,red) as a function of phase difference between layers $\varphi$. All currents vanish when $\varphi = 0$. Inset of (b) shows the dependence of the photocurrent at a representative frequency, $\omega_m = 2.5 t, \varphi = \pi/4$ as a function of tilting angle $\phi$. Lines are guides to the eye, following the function $\textrm{sin}(2\phi)$. All currents are normalized relative to the unit of current injection $j_0$ (see main text). 
    Here we retain all rotational symmetries such that TRS breaking is the only relevant perturbation (see Tab.~\ref{tab:1}). 
    }
    \label{fig:fig2}
\end{figure}

We note that the expression in Eq.~\eqref{eq:jqtensor} depends on the renormalized quasiparticle velocity $\mathcal{D}^a_{he} = \left[\sqrt{1-\Delta_k^2/\xi_e^2} \mathbf{v}^a_e - \sqrt{1-\Delta_k^2/\xi_h^2} \mathbf{v}^a_h\right]$ such that $\xi_{e/h}$ are the eigenenergies of the BdG Hamiltonians 
with velocities $v^a_{e/h}$
in the quasielectron/quasihole sectors, respectively. 
For a single-band and single-layer superconductor, $v^a_{e} = v^a_{h}$ and $\mathcal{D}^a_{he}$ vanishes identically, consistent with the optical absorption of clean superconductors \cite{Ahn_2021}. We decompose the current into specific quantum-geometric quantities, that couple differently to the incoming light polarization. The photocurrent for linearly polarized light reads,
\begin{align}
     \frac{\partial j^a_{\textrm{lin}}}{\partial t} = -\frac{ e^3 E_\alpha^2}{2\pi\hbar} \int \mathrm{d}^2 k \delta(\hbar\omega_0 - \Delta_{he})\mathcal{D}^a_{he}g^{\alpha \alpha}_{h e},
     \label{eq:jlin}
\end{align}
while for circular-polarized light we have,
\begin{align}
\frac{\partial j^a_{\textrm{circ}}}{\partial t} = \frac{ e^3 E_\alpha E_\beta^{*}}{4\pi\hbar} \int \mathrm{d}^2 k \delta(\hbar \omega_0 - \Delta_{he})\mathcal{D}^a_{he}\Omega^{\alpha \beta}_{h e}.
    \label{eq:jcirc}
\end{align}
The photocurrent arises from resonances between incoming light and the energy differences between occupied and empty BdG bands, i.e., $\Delta_{eh} = \xi_e-\xi_h$.
%i.e. $\Delta_{he} = 2\sqrt{\varepsilon^2_\mathbf{k}+\Delta_{\mathbf{k}}^2}$, where $\varepsilon_\mathbf{k}$ is the dispersion of the normal state. 
In addition to the photocurrent, Eqs.~\eqref{eq:jlin}-\eqref{eq:jcirc} , an out-of-plane polarization will be generated in a multilayer system by circularly polarized light. The rate is given by,
\begin{align}
    \dot{P}^z = \frac{ e^3 E_\alpha E_\beta^{*}}{4\pi\hbar} \sum_{l}\int \mathrm{d}^2 k \delta(\hbar \omega_0 - \Delta_{he})\chi^z_{l,he}\Omega^{bc}_{h e}.
    \label{eq:pz}
\end{align}
Here, we introduced the layer polarization operator for 2D BdG states: $\chi_{l,he}^z = \frac{d}{2} (\left \langle e| \mathcal{P}_{l} | e \right\rangle- \left \langle h| \mathcal{P}_{l} | h \right\rangle)$, where $\mathcal{P}_l$ is a projection operator onto layer $l$, and $d$ is the equilibrium interlayer separation. While in these matrix elements, a summation over all hole (occupied) and particle (empty) states is implied, in practice however, the frequency of light resonantly selects one pair of bands. In this sense, the nonlinear response is a direct probe of the quantum geometry of the BdG states. % In addition, there are contributions of lower order in $\Omega$ terms which do not lead to current injection. These are subleading and have zero efficiency in rectification. They are discussed in the SM and in the Discussion.
\paragraph{Off-normal response ---}
For 2D materials, confinement introduces a localized degree of freedom reproducing charge polarization. This permits a new class of optical response outside of the usual normal-incident description of optical conductivity \cite{Sipe2000}. For simplicity consider the case of light propagating on the $y-z$ plane, with its momentum $\mathbf{k} = |k| (\sin(\phi)\hat{y} + \cos(\phi)\hat{z})$. $\phi = 0$, see Fig.~\ref{fig:fig1}(a). This choice also dictates the polarization, such that $\mathbf{E}=\mathbf{E}
_0(-\cos(\phi),0,\sin(\phi))$ which is a TM mode.
%corresponds to the normal-incidence case. 
The perpendicular component (i.e., along $z$) couples to the charges through the dipole operator $\hat{p}_z = -e\hat{z}$ (see SM). 
We find a new contribution that is allowed for all chiral point groups (i.e., lacking all mirror symmetries), which is naturally realized in twisted homobilayer, and stacked (or twisted) heterobilayers. 
The chiral response -- which arises when all mirror symmetries are broken and is non-vanishing for all rotational groups is -- 
\begin{align}
    \frac{\partial j_{\textrm{chiral}}}{\partial t}= -\frac{e^3 E^2}{4\pi \hbar}\sin(2\phi) \int \mathrm{d}^2 k \delta(\hbar \omega_0 - \Delta_{he})\mathcal{D}^{\perp}_{he}Q^{\parallel}_{h e}, 
    \label{eq:chiral_1}
\end{align}
For the generality of the expression, we adopted the following notation: $\perp$ denotes the in-plane direction perpendicular to the in-plane component of the polarization (i.e., $x$, if the polarization is in the $y-z$ plane). $\parallel$ denotes the components defined in the plane of polarization, that is $Q^{\parallel} = (\mathcal{A}^y p^z)_{he}$. Similarly to the decomposition above, we may separate the current response into its quantum metric, $g^{\parallel} = 2 \textrm{Re}(Q^{\parallel})$, and Berry curvature parts $\Omega^{\parallel} =  \textrm{Im}(Q^{\parallel})$. This response is shown in Fig.~\ref{fig:fig2} for TBSCs and described in detail below.
\allowdisplaybreaks

\paragraph{Symmetries ---} 
We now analyze the point-group, TRS, and particle-hole symmetry constraints on Eqs.~\eqref{eq:jlin}-\eqref{eq:pz}. 
As current is mediated by a rank-2 response tensor, i.e., $j^a = \chi^{a;\alpha \beta} E_\alpha E_\beta$, inversion symmetry ($P$) must be broken. We focus on cases in which inversion is broken in the normal state, that is $PH_0(\mathbf{k})P \neq H_0(-k)$, but TRS is otherwise preserved. This condition however is insufficient in order to obtain a finite current in the superconducting state. For simplicity, we disregard contributions from transitions that are allowed in the normal (non superconducting i.e. ``unfolded'') state \cite{Watanbe2022nonreciprocal}. This is justified as with $\Delta_\mathbf{k} = 0$, there are no resonances at the frequencies of interest. 
%\jp{JP: Why is this justified?}

The momentum dependence of the pairing term then determines the condition for a non-zero current. As is shown below, all rotational symmetries suppress some components of the in-plane current. We represent the symmetry operation via a matrix $R_{\alpha'\alpha}$ such that under the symmetry operation, $\chi^{a; \alpha \beta} = 
R_{aa'}R_{\alpha \alpha'}R_{\beta \beta'} \chi^{a';\alpha'\beta'}$. 

%\jp{JP: Why is the order the opposite on the $a$ index?}

We note two important symmetries: mirror and TRS. In-plane mirror symmetries rotate the spin while accomplishing an incomplete rotation of the particle-hole basis, $\psi_\mathbf{k} = (c_{\mathbf{k}, \uparrow},c^\dagger_{-\mathbf{k}, \downarrow}) \to (c_{M_{x/y} \mathbf{k}, \downarrow},c^\dagger_{M_{x,y} \mathbf{k}, \uparrow})$. Thus along any mirror line, there exists an exact particle hole symmetry which causes any longitudinal current, i.e. $j \parallel E$, to vanish. Also important is the action of TRS. For TRS, the particle-hole basis is rotated as $\psi_\mathbf{k} = (c_{\mathbf{k}, \uparrow},c^\dagger_{-\mathbf{k}, \downarrow}) \to i \sigma_y \psi_{-\mathbf{k}}^{\dagger}$.
This again would lead to the vanishing of quantities (such as $j_\textrm{lin.}$), as they are odd under this basis rotation. Concretely, as $\textrm{Re}Q^{\alpha \beta}$ in Eq.~\eqref{eq:jqtensor} is both particle-hole and TRS even, the vanishing of the quantum metric current depends on a breaking of TRS in \textit{either} the velocity renormalization factor $\mathcal{D}_{he}$ or the BdG spinor for a given layer, $\Psi_\mathbf{k}^\dagger = \left[c^\dagger_{\mathbf{k
}, \uparrow},c_{-\mathbf{k}, \downarrow} \right]^T$.
% {\bf JP: $\Psi$ where is this defined?}
If TRS is preserved (in the normal velocity $\mathbf{v}$ \textit{and} the pairing kernel $\Delta_\mathbf{k}$) any currents proportional to the quantum metric vanish. We fully analyzed the symmetry structure for all symmetries of 2D point groups (where the operations do not mix particles and holes) for the response tensor $\chi^{a;\alpha \beta}$ in Tab.~\ref{tab:1} and decomposed it to its constituents: response to linear polarized light ($\propto g^{\alpha \beta}$) and circularly polarized light, ($\propto \Omega^{\alpha \beta}$). 
A non-vanishing current strongly depends on mirror symmetries that cause reflection in the plane. As mirror operations also rotate the spin, they would exchange spin indices and the definition of particles and holes. As a result,  mirror symmetries cause certain components of the current to vanish, in the direction \textit{parallel} to the light mirror plane, see Tab.~\ref{tab:1}.

\begin{table}[h]
\caption{Symmetry restrictions on the in-plane, out-of-plane, and chiral current. If a given symmetry is present we mark an allowed (disallowed) response with $\checkmark$ ($\xmark$).
}
\label{tab:1}
\begin{tabular}{|c|c|c|c|c|}%{|p{.2\linewidth}|p{.2\linewidth} |p{.2\linewidth}|p{.22\linewidth}|p{.20\linewidth}|}
\hline
Symmetry & $j_{\textrm{lin}}$ & $j_{\textrm{circ}}$ & $P_z$ & $j_\textrm{chiral}$\\ \hline 
  $C_{2z}$  & $\xmark$ & $\xmark$ & $\checkmark$ & $\checkmark$\\ 
  $C_{3z}$  & $\checkmark$ & $\xmark$ & $\checkmark$ & $\checkmark$ \\ 
  $C_{4z}, C_{6z}$  & $\xmark$ & $\xmark$ & $\checkmark$ & $\checkmark$ \\
  $C_{2x}$  & $\checkmark$, $j_y = 0$ & $\xmark$ & $\checkmark$ & $\checkmark$ \\ 
  $M_{x/y}$  & $\checkmark$, $j \parallel M_{x/y} = 0$ & $\xmark$ & $\xmark$ & $\xmark$   \\ 
  $\textrm{TRS}$  & $\xmark$ & $\checkmark$ & $\xmark$ & $\textrm{circ.}~ \checkmark, \textrm{lin.}~\xmark$ \\
  \hline
  
\end{tabular}
\end{table}

\paragraph{Photocurrents and polarization in topological and nodal superconductors ---}

We now apply the theory to twisted nodal superconductors, using the continuum model of Ref.~\cite{Volkov2023magic} in the small twist angle limit, with flakes of the high temperature superconductor Bi\textsubscript{2}Sr\textsubscript{2}CaCu\textsubscript{2}O\textsubscript{8+$\delta$} (Bi-2212) \cite{Can_2021, zhao2023time} serving as the experimental setting. 

Near a nodal point $K_N$, the minimal model takes the form, $H=H_N+H_{\textrm{non-circ.}} + \delta H_{\textrm{sym.}}$ where the low energy twisted bilayer continuum model is
\allowdisplaybreaks
\begin{equation}
    \notag H_N = \sum_{\mathbf{k}} \Psi^\dagger_\mathbf{k}\bigl[v_F k_{x} \tau_3 \sigma_0 + v_\Delta k_y \tau_1\sigma_0  -\frac{v_\Delta Q_N}{2}\tau_1  \sigma_3 + t \tau_3 \sigma_1  \bigr]\Psi_\mathbf{k} .
    \label{eq:Ham}
\end{equation}
Here, momentum ${\bf k}$ is expanded near the node momentum with $k_{x/y}$ labeling the local orthogonal coordinate system. $\Psi_\mathbf{k}$ is the Nambu spinor which includes the layer index with the appropriate momentum shift due to the twist, i.e., $\Psi_{\mathbf{k}} = \left[\Psi_1(\mathbf{k}-\mathbf{Q}_N/2),\Psi_2(\mathbf{k}+\mathbf{Q}_N/2)\right]$, such that $1,2$ are the layer indices. The Hamiltonian is represented by matrices $\tau_i$ and $\sigma_i$ acting in Gor'kov-Nambu and layer space, respectively, and $\mathbf{Q}_N$ is the twist induced momentum transfer $\mathbf{Q}_N = \theta(\hat{z} \times \mathbf{K}_N)$ (for small twists). The parameters $v_F$ and $v_\perp$ are the Fermi velocities of the electrons and the linear dispersion of the gap function, respectively, and $t$ is the tunneling strength between the layers at a node $K_N$. 
For a circular Fermi surface there is a ``magic" value of the twist $\theta_{MA} = 2 t / (v_\Delta K_N)$ where the velocity of the BdG Dirac cones vanishes realizing a quadratic band touching \cite{Volkov2023magic}. Without non-circular corrections, the photocurrent is zero due to a continuous rotational symmetry. The relevant contributions from a typical Bi-2212 Fermi surface are then $H_{\textrm{non-circ.}}= v_F^{(2)}k_y (\theta/2) \tau_3 \sigma_3 - v_\Delta^{(2)}k_x (\theta/2) \tau_1 \sigma_3 $ where $v_F^{(2)}\sim v_F $ and $v_\Delta^{(2)}\sim v_{\Delta}$ are the curvatures of $v_{F/\Delta}$, and are required for the photocurrent to not vanish. The model for twisted Bi-2212 retains a combination of $C_{4z}, C_{2x}$ which causes normal-incidient photocurrents to vanish, see Tab.~\ref{tab:1}. We add terms $\delta H_{\textrm{sym.}}$ to break this symmetry (specifically corresponding to a supercurrent biasing two nodes breaking $C_{4z}$, $C_{2z}$) leading to a non-vanishing current. Concretely, the full Hamiltonian comprising all nodes is $H = \textrm{diag}(H_\mathbf{k},H_{C_{4z}\mathbf{k}},H_{C_{4z}^2\mathbf{k}},H_{C_{4z}^3\mathbf{k}}) + (0,-\hbar vq_f\tau_0,\hbar vq_f \tau_0,0)$.
We choose $\hbar vq_f = 0.1t$ for the symmetry breaking. The added Cooper pair momentum $vq_f$ acts as $\tau_0$ in the Nambu basis for each node. 

The sensitivity of the photocurrent to TRS allows us to study the order parameter structure. In the limit of small twist angles, application of an interlayer current (that breaks TRS) induces a topological $d+id$-like superconducting state~\cite{Volkov2023current,Volkov2023magic}, whereas the interactions between quasiparticles at $\theta\approx\theta_{MA}$ could spontaneously break the TRS leading to, e.g. $d+is$ superconducting state~\cite{Volkov2023magic,Tummuru-2022}.
For the $d+id$ case, effect of the interlayer current is to ``rotate'' the particle-hole basis in the form:
    $\tau_1 \rightarrow \cos\frac{\varphi}{2} \tau_1 - \sin\frac{\varphi}{2} \tau_2\sigma_3.$ 
For $d+is$, the $is$ pairing includes a term $\psi_0 \tau_2$.

In Figs.~\ref{fig:fig2}(a)-(b) we plot the chiral photocurrent injection rate separating the linear polarization $j_\textrm{lin}$ and circular polarization $j_\textrm{circ.}$, without additional symmetry breaking setting $\delta H_{\textrm{sym}} = 0$. We obtain the full response by adding contributions of four symmetry-related Dirac node pairs. All currents are plotted with respect to the natural unit of current injection in this model, $j_0 = \frac{e^3 \sqrt{v_f v^{(2)}_f}}{4\pi \hbar t} E^2$. For typical values $E \sim 1~\textrm{mVnm}^{-1}$, $t \sim 1~\textrm{meV}$, $v_f = 10^{5}~\textrm{m/s}, v^{(2)}_f \approx v_f/2$, giving an injection rate of $j_0 \sim 1.3\cdot 10^{6}\frac{\textrm{A}}{\textrm{nm} \cdot \textrm{sec}}$. We stress that this current probes a piece of the usual geometric quantities $g^{\alpha \beta}, \Omega^{\alpha \beta}$, and their projection on the light polarization plane.
For a pulse of duration $\Delta \tau \sim 10^{-9}~\textrm{sec}$ we expect a 2D current density of $\sim \frac{1 \mu A}{\textrm{nm}}$, where we include the rough scale of $j/j_0 \sim 10^{-3}$. This current magnitude should be easily detected using available experimental techniques. 

The photocurrent is nonvanishing above $\omega \gtrsim 2t$ ($t$ is the interlayer tunneling at a node) which shows that the current is dependent on switching the layer degree of freedom, in agreement with the conventional optical response of superconductors, which prevents transition between exact particle-hole symmetric bands \cite{Ahn_2021}. In Fig.~\ref{fig:fig2}, we show the evolution of the off-normal photocurrent with twist angle $\theta/\theta_{\textrm{MA}}$ for linearly polarized light (probing the quantum metric) in (a) and circularly polarized light (probing the Berry curvature) (b) with the former achieving a maximum at $\theta= \theta_{MA}$, and their values are directly tunable by varying the interlayer current (through the interlayer phase difference $\varphi$) in  Fig.~\ref{fig:fig2}(a;inset). Similarly, we find that for the  TRS broken state near twists of $\theta = 45^{\circ}$, these off normal chiral responses are also non-zero (not shown).

When the symmetry of the continuum model is broken down further by biasing a pair of nodes (thus removing $C_{4z}, C_{2z}$), a finite photocurrent in the normal incidence geometry emerges. 
 \begin{figure}
%     \centering
     \includegraphics[width=0.95\columnwidth]{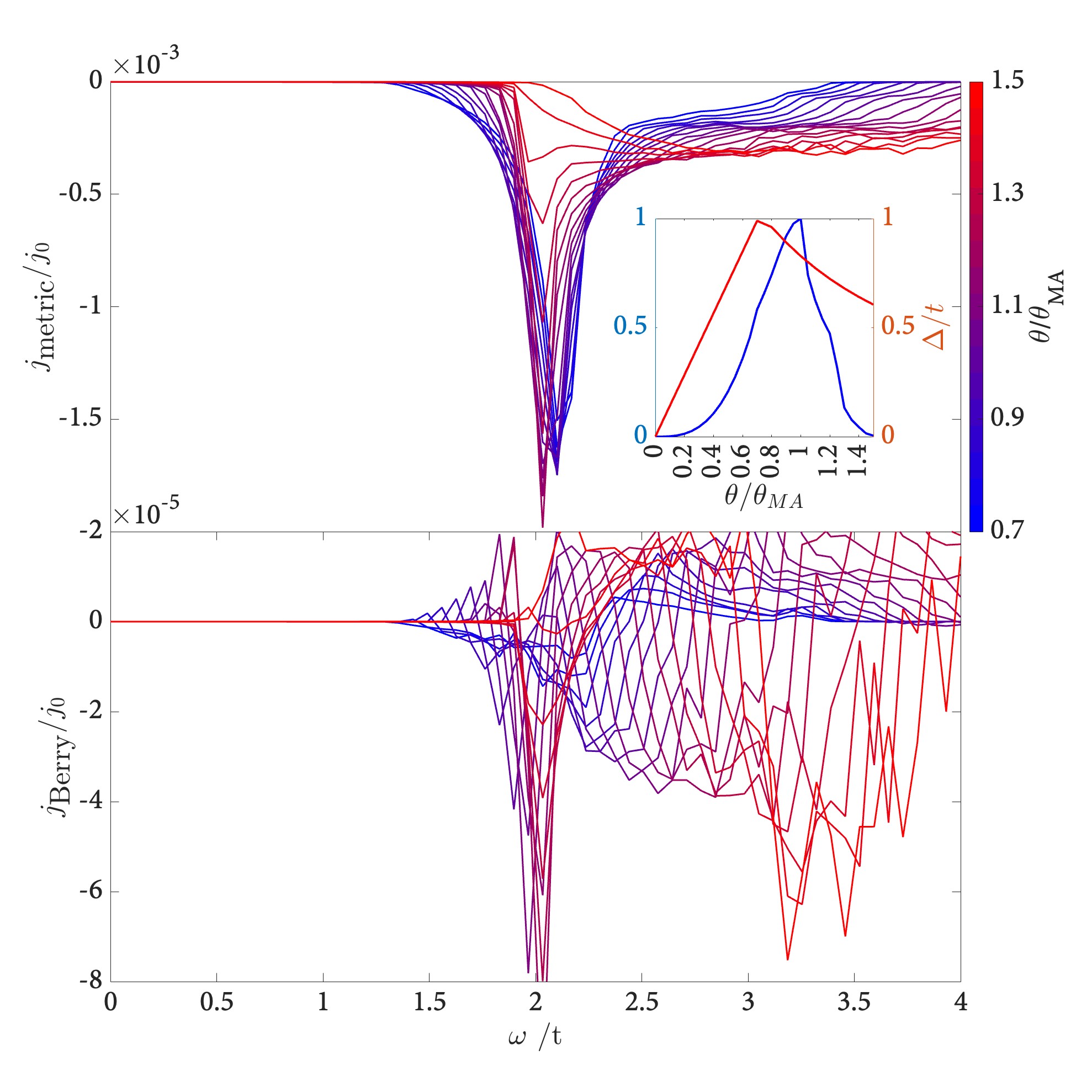}
    \caption{{\bf Normal incidence (in-plane) photocurrent for TBSCs}: Here we focus on the Hamiltonian in Eq.~\eqref{eq:Ham} with additional symmetry breaking (see SM). Top: Linear polarized light response, related to the quantum metric $g^{\alpha \alpha}$. bottom: Circular polarized light response, proportional to $\Omega^{xy}$, the quasiparticle Berry curvature. Inset of top shows the dependence of the current (blue line) on the twist angle, which is correlated with the onset of the maximal gap $\Delta/t$, as in Ref.~\cite{Volkov2023current}.
    }
     \label{fig:fig3}
 \end{figure}
We plot the normal incident ($\phi = 0$) photocurrent components in Fig.~\ref{fig:fig3}(a)-(b). With the additional symmetry breaking and layer-particle mixing, a finite signal appears for a finite frequency near $2t$. This is expected for a system with reduced symmetry. Both response components are sharply peaked at $\omega \sim 2t$ due to the emergence of a band edge at this value. It should be noted that the maximum value of photocurrent is obtained at a frequency of $\approx 2t \pm \Delta$, where $\Delta$ is the topological gap. The non-vanishing optical matrix elements are separated by $2t$ while the topological states are $\pm \Delta$ within that manifold. In this context, we note that although the chiral current is non-vanishing for any rotational group, it can be used to measure the rotational symmetry; in Fig.~\ref{fig:fig4} we plot the dependence of the current $j^x$ as a function of the rotation of the polarization angle $\phi$. For the non-symmetry broken case, Fig.~\ref{fig:fig4}(a) shows a four-fold symmetric current pattern. When the rotational symmetry is broken, as in  Fig.~\ref{fig:fig4}(b), the current is anisotropic and two-fold invariant, clearly heralding the reduced symmetry.

\textit{Applications to tWSe$_2$ and RMG}: We now apply our theory directly to several recently observed moire superconductors. This allows us to provide concrete predictions that can directly test for the symmetry of the superconducting state. We begin with tWSe$_2$, which has been observed to superconduct in near $3^\circ$ and 5$^\circ$ twisted bilayers \cite{xia2024superconductivity,guo2025superconductivity}. 
Recent theoretical work \cite{guerci2024topologicalsuperconductivityrepulsiveinteractions,Wietek2022,fischer2024theoryintervalleycoherentafmorder,schrade2024nematicchiraltopologicalsuperconductivity}, has narrowed down the most likely pairing symmetry to an admixture of spin singlet and triplet (due to broken inversion from the displacement field) and either a nematic (i.e. nodal) or chiral (i.e. gapped) superconducting ground state though the precise lowest energy configuration depends sensitively on the pairing mechanism. In the chiral state a response to normal-incident circularly polarized light is generically forbidden, though as it has broken TRS a non-zero $j_\textrm{chiral}$ will appear under linear polarized illumination. In addition, as the relevant rotational symmetry is $C_{3z}$, a normal incidence photocurrent response for linearly-polarized light will be allowed once TRS is broken.
In contrast, for the nematic state (preserving TRS, but breaking $C_{3z}$ symmetry, with $C_{2x}$ already broken by displacement field), a finite response to circularly polarized light will appear for normal incidence. For chiral state, this response will remain zero due to preserved $C_{3z}$.

For this response we estimate it appears on the frequency scale of $\omega \sim 2t \sim 50 \textrm{meV}$ where $t$ is the scale of interlayer tunneling \cite{devakul2021magic}. 

Recent experiments \cite{han2024signatureschiralsuperconductivityrhombohedral,yang2024diverseimpactsspinorbitcoupling} on multilayer rhombohedral graphene structures show superconductivity emerging at finite displacement field for several doping regions, in close proximity to symmetry broken states, showing, e.g., an anomalous Hall effect. It is unclear whether the superconducting state breaks TRS and whether the order parameter breaks additional crystal symmetries (for example, through nematicity). We propose using the photocurrent for the tomography of the order parameter: the finite displacement field breaks inversion symmetry in the system without breaking other crystal symmetries (mirrors or rotation in the plane). It will then be possible to probe rotational symmetry breaking by the order parameter, if a response to normal-incident circularly polarized light is detected.

\begin{figure}
    \centering
    \includegraphics[width=0.95\columnwidth]{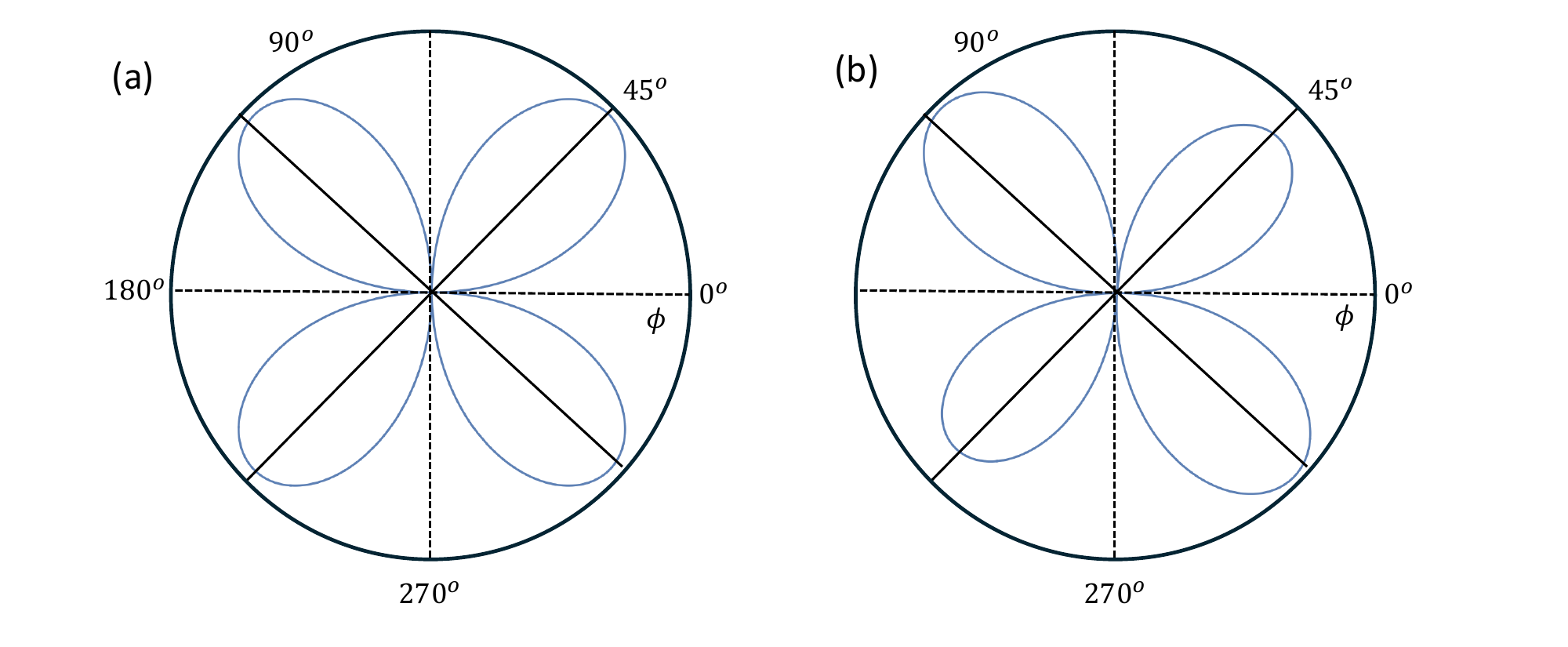}
    \caption{{\bf Detecting symmetry breaking with the chiral current.} (a) Chiral current in the $C_{4z}$ symmetric case, corresponding to $j^x/\textrm{max}(j^x)$ as a function of polarization angle $\phi$. A four-fold pattern emerges, due to the four-fold symmetry of the model. (b) When the symmetry is broken, e.g., by a in-plane current, the symmetry reduces to two-fold, as shown. This will be a clear signature of for spontaneous symmetry breaking, as well. }
    \label{fig:fig4}
\end{figure}

\textit{Discussion}-- In summary, we have presented a theory for photocurrents in topological nodal superconductors. We derived an expression for the photocurrent which depends on the quantum geometry of BdG bands: the momentum-space dipoles of the quantum metric and the Berry curvature. Both are sensitive to the existence of a topological gap, but the linear polarized current, in particular, directly probes whether TRS is broken in the ground state. We comment here on additional, low order terms disregarded here: firstly, a shift current may appear but depends on a non-vanishing $\partial_{k}^2 H$ \cite{Xu_2019} vertex which does not exist in the studied model; in addition, these terms vanish in the clean limit \cite{campos2023intrinsic}. The injection type of response scales as $1/\Omega$, while a shift current scales as $\Omega^{0}$. Therefore, combining with the absorption (optical conductivity) $\alpha \sim 1/\Omega$ the injection response $j/\alpha \sim \textrm{const.}, \Omega \to 0$, while for the shift current $\sim \Omega \to 0$, as $\Omega \to 0$. Thus, our calculated photocurrent is the relevant one in the clean limit for rectification.
We have additionally presented the use of the photocurrent to map the magic angle in twisted systems, and the tomography of order parameters in novel superconductors with broken inversion symmetry such as twisted TMDs and multilayer graphenes with an out of plane displacement field. We expect the photocurrent maybe integrated into ``on-chip" setups as recently shown in Ref.~\cite{seo2024chip}.

\paragraph{Acknowledgements---}
We thank Marcel Franz,  Philip Kim, and Matteo Mitrano for useful discussions.
D. K. is supported by the Abrahams postdoctoral Fellowship of the Center for Materials Theory, Rutgers University, and the Zuckerman STEM fellowship. K.P.L. and J.H.P. are partially supported by NSF Career Grant No.~DMR-1941569 and the Alfred P. Sloan Foundation through a Sloan Research Fellowship. This work was  performed  in part at the Aspen Center for Physics, which is supported by National Science Foundation grant PHY-2210452 (D.K., P.V., and J.H.P.). 
All authors acknowledge partial support by grant NSF PHY-2309135 to the Kavli Institute for Theoretical Physics (KITP).
\bibliography{main_v2.bbl}
\clearpage
\onecolumngrid
\begin{center}
{\bf{Suppelemental Material: Quantum geometric photocurrents of quasiparticles in superconductors}}
\end{center}
\setcounter{equation}{0}
\setcounter{figure}{0}

\renewcommand\theequation{S\arabic{equation}}
\renewcommand\thefigure{S\arabic{figure}}

% \author{Daniel Kaplan}
% \affiliation{%
%  Department of Physics and Astronomy, Center for Materials Theory,
% Rutgers University, Piscataway, NJ 08854, USA
% }%
% \author{Kevin P.~Lucht}
% \affiliation{%
%  Department of Physics and Astronomy, Center for Materials Theory,
% Rutgers University, Piscataway, NJ 08854, USA
% }%
% \author{Pavel A.~Volkov}%
% \affiliation{%
% Department of Physics, University of Connecticut, Storrs, Connecticut 06269, USA
% }%
% \author{J.~H.~Pixley}
% \affiliation{%
%  Department of Physics and Astronomy, Center for Materials Theory,
% Rutgers University, Piscataway, NJ 08854, USA
% }%
% \affiliation{
% Center for Computational Quantum Physics, Flatiron Institute, 162 5th Avenue, New York, NY 10010
% }%

%\maketitle
Here, we present the derivation of photoconductivity, the compact expressions for the current presented in the main text and the discussion of the symmetries of the model used in the calculation in the main text.
\section{General theory of photoresponse}
The perturbation we consider involves an electric field $\mathbf{E}(t) = -\partial_t \mathbf{A}(t)$ produced by coherent (i.e. laser) light represented by a wave at frequency $\omega_0$ with a slight detuning $\Omega$. This two photon process can be represented by a two color signal,
\begin{align}
    \mathbf{A}(t) = \frac{1}{2}\mathbf{A}_0 \left(\cos(\omega_0 t)+\cos((\omega_0 +\Omega) t)\right),
\end{align}
where we shall be interested in the $\Omega \to 0$ limit.

This signal has the Fourier representation,
\begin{align}
    \mathbf{A}(\omega) = \frac{1}{2}\mathbf{A}_0 \left(\sum_{p=\pm}\delta(\omega + p\omega_0)+\delta(\omega + p(\omega_0 +\Omega))\right)
\end{align}
This expression essentially describes two signals whose frequency is mismatched by a small amount, $\Omega$. In the limit of $\Omega \to 0$ we recover the usual case of continuous wave illumination. We calculate the Green's function $G(\mathbf{E},\mathbf{k})$ in accordance with the Dyson equation (For a similar discussion on perturbative solution to Dyson's equations, see Ref. \cite{Xu2022nonlinear}). In general, we seek all contributions for which $\omega_1+\omega_2 = \pm \Omega$ which results from energy conservation. We disregard other contributions such as ($2\omega_0 \pm \Omega$) contributions as they do not lead to a strongly enhanced response, as we demonstrate below. In the leading order in $\Omega \to 0$, the currents derived (and their matrix elements) are related to the photovoltaic effect derived for charge currents \cite{kaplan2023unifying,holder2020consequences,Parker2018,Xu_2019}.
The central question is the calculation of an expectation value for a vector operator, $O_a$, and it is defined by:
\begin{align}
    \left\langle \hat{O}_a \right\rangle = -\frac{i}{(2\pi)^3} \int \mathrm{d}E \mathrm{d}^2k  \textrm{Tr}\left(\hat{O}_a G(E,\bff{k})\right).
    \label{eq:result}
\end{align}
% \jp{JP: This is missing the bold on the dk and a factor of $1/(2\pi)^2$ to match the main text.}
Here $\hat{O}_a$ is either the polarization operator $p_z$ or the current vertices $v^{x}, v^y$. $G(E,\mathbf{k})$ is the full Green's function at momentum $\mathbf{k}$ and frequency $E$.
\subsection{Difference frequency photovoltaic effect}
The Green's function in the presence of the EM field can be calculated perturbatively from the Dyson equation. The equation is,
\begin{align}
    G = G_0 + G_0\Sigma G,
    \label{eq:Dyson}
\end{align}
which is solved iteratively, starting with $G = G_0$. $G_0$ is the effective, $T=0$, Green's function in the BdG eigenbasis. Here, $\Sigma$ is the self-energy, which in the present case arises due to the coupling to the EM field.
We note that $\Sigma$ as derived from minimal coupling has no off-diagonal terms in the Nambu basis, as it is represented by $\tau_0$.

The general solution to Eq.~\eqref{eq:Dyson} \cite{jishi2013feynman,Mahan2000} can be written in series form. Working in a basis where we use the retarded $G_r$, advanced, $G_a$, and equilibrium $G_0$ Green's functions, the perturbtive solution to the full $G$ takes the form,
\begin{align}
    G(E) = \sum_{l_1 =0 }^{\infty} (G_r \Sigma)^{l_1} G_0 \sum_{l_2=0}^{\infty} (G_a \Sigma)^{l_2}.
    \label{eq:greenfull}
\end{align}
\subsubsection{The self energy}
We now define the interaction with the field, encoded through the self-energy. The latter is obtained by a series expansion of the BdG Hamiltonian. We define,
\begin{align}
    H_{\textrm{BdG}}(\mathbf{k},\mathbf{A}) = \left(\begin{matrix}H_0 (k-e\mathbf{A}) & \Delta_\mathbf{k} \\ \Delta^\dagger_{\mathbf{k}} & -H_0 (-k-e\mathbf{A}) \end{matrix}\right)= H_{\textrm{BdG}}(\mathbf{k}, 0) + \underbrace{\left(\begin{matrix} \partial_{k}H_0 (k) & 0 \\ 0 & \partial_{k} H_0 (-k) \end{matrix}\right)}_{\mathbf{J}_{\textrm{BdG}}} \mathbf{A} = H_\textrm{BdG} + \Sigma.
\end{align}
This leads us to identify, $\Sigma = \tau_0 \partial_k H_0$.
\subsubsection{Polarization operator}
\label{sec:polarization}
We follow Ref.~\cite{gao2020tunable} in defining the layer polarization operator for few-layer systems. In the normal state, the wavefunction may be assumed to be exponentially localized $|\psi(z)|^2 \sim e^{-z^2/d_0^2}$. Thus, the projection on the two layers may be defined as,
\begin{align}
p_z = (e d/2)\sigma_3,
\end{align}
with $\sigma_3$ ensuring that the charge polarization flips sign between layers. $d$ is  the distance between layers. In the superconducting state, the Nambu components must flip as the coupling to the electric field is odd under charge conjugation. Thus, $\hat{p}_z = (e d/2) \sigma_3 \tau_3$. This expression can be deduced from the following representation in the Nambu basis. Without loss of generality, focusing on one layer, the Nambu spinor takes the form $\Psi_\mathbf{k}^\dagger = \left[c^\dagger_{\mathbf{k
}, \uparrow},c_{-\mathbf{k}, \downarrow} \right]^T$, the free energy in the presence of finite polarization may be written as, 
\begin{align}
F = F(p^{+}=0) + \sum_{\mathbf{k}} p^{+}_\mathbf{k} + \Psi^\dagger_{\mathbf{k}}\left(\begin{matrix}
p^{+}_\mathbf{k} & 0 \\ 0 & - p^{+}_{\mathbf{k}}
\end{matrix}\right)\Psi_{\mathbf{k}} = F(p^{+} = 0) + \sum_\mathbf{k} p_\mathbf{k}^{+}\left(c^\dagger_{\mathbf{k},\uparrow}c_{\mathbf{k},\uparrow} + c^\dagger_{\mathbf{k},\downarrow}c_{\mathbf{k},\downarrow}\right),
\label{eq:free_e}
\end{align}
with $p^{+}_{\mathbf{k}}$ being the projection of Bloch state onto the top layer $(+)$, and Eq.~\eqref{eq:free_e} recovers the total polarization in the layer, correctly. The expression of the opposite layer involves flipping the form of $p^{+}$ to $p^{-}$, which amounts to changing the sign of the term, assuming both layers are separated by a distance $\pm d/2$. Thus, the polarization operator is $\hat{p}_z = (ed/2)\sigma_z \tau_z$. The associated self-energy in the perturbative series is then $\Sigma = p_z E_z$.
\subsubsection{Dyson series}
We expand the Green's function of Eq.~\eqref{eq:greenfull} to second order in $\mathbf{A}$ (which is identical to second order in $\mathbf{E}$). In the series, this is contributed by the terms with $l_1 = 2, l_2 = 0$, $l_1 = 1, l_2 = 1$ and $l_1 =0, l_2 =2$. The net energy in the fermion loop is kept to be $\Omega$, as that is the probing frequency of the current. The injection current will be the term diverging in the limit $\Omega \to 0$. Thus, the probing frequency $\Omega \ll 2\Delta$ but larger than the typical disorder strength (relevant for low energies in superconductors, allowing us to take the clean limit. The expansion is carried out for generic field frequencies $\omega_1,\omega_2$ for generality. At the end of the calculation, we substitute $\omega_1 = \omega_0, \omega_2 = -\omega_0 +\Omega$ and symmetrize, since the order $\omega_1, \omega_2 $ of insertions is arbitrary. Thus,
\begin{align}
    \notag G(E,\bff{k}) &=  \frac{1}{2}G_r^E \Sigma(\omega_1) G_r^{E+\omega_1}\Sigma(\omega_2) G_0^{E+\omega_1+\omega_2}+G_0^E \Sigma(\omega_1) G_a^{E+\omega_1}\Sigma(\omega_2) G_a^{E+\omega_1+\omega_2} + G_r^E \Sigma(\omega_1) G_0^{E+\omega_1}\Sigma(\omega_2) G_a^{E+\omega_1+\omega_2} + (\omega_1 \Leftrightarrow \omega_2).
\end{align}
As usual, $G_r(\omega_1, \mathbf{k}) = \frac{1}{i\omega_1 -H_\textrm{BdG}(\mathbf{k})+i\Gamma}, G_a = \frac{1}{i\omega_1 -H_\textrm{BdG}(\mathbf{k})-i\Gamma}$, $G_0(\omega_1,\mathbf{k}) = G_R - G_A$, and we are using the shorthand $G_0^{\omega_1}\equiv G_0(\omega_1,\mathbf{k})$.
The techniques for evaluating the integral over $\textrm{d}E$ are listed in Refs.~\cite{Mahan2000,Parker2018,Watanbe2022nonreciprocal,Michishita2021,kaplan2023general,kaplan2023general}.
We decompose the quantum average of Eq.~\ref{eq:result} into three constituent pieces. The thermodynamic trace is evaluated by inserting the complete set of states for the BdG Hamiltonian and evaluating $G$ in its Lehmann (spectral) representation,
\begin{align}
    \textrm{Tr}\left[O^a G_r v^\alpha G_r v^\beta G_0 \right] = \frac{(2\pi i f_m) O^a_{mn} v^\alpha_{nl} v^\beta_{lm}}{(\varepsilon_m-\varepsilon_n-\hbar(\omega_1+\omega_2)+i\Gamma)(\varepsilon_m-\varepsilon_l -\hbar\omega_2+i\Gamma)} A^\alpha (\omega_1) A^\beta(\omega_2).
    \label{eq:part1}
\end{align}
\begin{align}
    \textrm{Tr}\left[O^a G_0 \Sigma(\omega_1) G_a \Sigma(\omega_2) G_a \right] = \frac{(2\pi i f_n) O^a_{mn} v^\alpha_{nl} v^\beta_{lm}}{(\varepsilon_n-\varepsilon_l +\hbar\omega_1-i\Gamma)(\varepsilon_n-\varepsilon_m +\hbar(\omega_1+\omega_2)-i\Gamma)}A^\alpha (\omega_1) A^\beta(\omega_2).
\end{align}
\begin{align}
    \textrm{Tr}\left[O^a G_r \Sigma(\omega_1) G_0\Sigma(\omega_2) G_a \right] = -\frac{(2\pi i f_l) O^a_{mn} v^\alpha_{nl} v^\beta_{lm}}{(\varepsilon_l - \varepsilon_n-\hbar\omega_1+i\Gamma)(-\varepsilon_l + \varepsilon_m+\hbar\omega_2+i\Gamma)} A^\alpha (\omega_1) A^\beta(\omega_2).
    \label{eq:part3}
\end{align}
$n,l,m$ run through the BdG states. $\varepsilon$ denotes the eigenvalues of $H_\textrm{BdG}$. $\omega_1, \omega_2$ are the frequencies of the perturbing fields. $f_{m}$ is the occupation factor for the quasiparticle band $m$ and momentum $\mathbf{k}$.  
These expressions are now to be simplified using the condition formulated above that $\omega_1 = \omega_0, \omega_2 = -\omega_0 - \Omega$. Note that by construction, we evaluate all Green's functions in the limit $\Gamma \to 0$ (clean limit). In the limit of $\Omega \ll 2\Delta$, it is sufficient to identify the most divergent part of these equations above. In Eqs.~\eqref{eq:part1}-~\eqref{eq:part3}, the divergent terms with power $1/\Omega$ are the only relevant contributions, as they lead to an operator expectation value growing in time. 
We stress again that the current is evaluated at the frequency $\Omega$ at which the Green's function also diverges as (at least) $\Omega^{-1}$. The result for the non-analyticity is found in the time domain. We recall that by the properties of Fourier transforms (FT) $i \Omega \langle O \rangle_{\Omega}\overset{\textrm{FT}}{\mapsto} \frac{\partial \langle O \rangle}{\partial t}$. The divergent terms are known as injection currents (when the calculated vertex is that of a current). In the case of polarization, we term this effect as interlayer charge injection. All the currents derived here are of the injection type, as these are the only ones capable of non-zero efficiency rectification in the clean limit \cite{campos2023intrinsic}.

% The combination of all three processes (for the different self energies $\partial_k H_0$ and $p_z$) can be represented diagrammatically: 

\subsection{Injection phenomena in superconductors}
In the above equations, the response is computed in real frequency (after analytic continuation). As we show below, the response of an operator $O$ generally takes the form, 
\begin{align}
    O^a = \frac{1}{i\Omega} \chi^{a,bc}(\Omega) A_b(\omega_0) A_c(-\omega_0+\Omega)
\end{align}
In the time domain, this corresponds to the following,
\begin{align}
    \frac{\partial O^a(t)}{\partial t} = 
    \int \textrm{d} \Omega e^{-i\Omega t}\chi^{a,bc}(\Omega)A_b(\omega_0)A_c(-\omega_0+\Omega).
\end{align}
We remark that the operator $O$ is an observable and hence purely real. Thus, it allows for the decomposition and temporal integration
\begin{align}
    O^{a}(t) = \int_0^{t} \textrm{d}t'\int \textrm{d}\Omega \left[\cos(\Omega t') \chi_1^{a,bc} + \sin(\Omega t')\chi_2^{a,bc}\right]A_b A_c.
\end{align}
Here, $\chi_1^{a,bc} = \textrm{Re}\left(\chi^{a,bc}\right), \chi_2^{a,bc} = \textrm{Im}\left(\chi^{a,bc}\right)$. For now, we assume that $A_{b,c}$ are purely real themselves.
For a very narrow pulse, $A_b(-\omega+\Omega)A_c(\omega)$ is peaked only when $\Omega \to 0$ or $\Omega \approx \pm 2\omega$. Thus, in what follows we assume that $A_b(\omega_0) A_c(-\omega_0 +\Omega) \approx \delta(\Omega)$. The generalization to a wider pulse is readily possible by keep the broadening finite and convolving the integral in its full form. However, as these contributions are not divergent in time, they can be disregarded.
\begin{align}
    O^{a}(t) = \lim_{\Omega \to 0} \left(\frac{\sin(\Omega t)}{\Omega} \chi_1^{a,bc} - \frac{(\cos(\Omega t)-1)}{\Omega}\chi_2^{a,bc}\right)A_b A_c = \chi_1^{a,bc}A_b A_c t.
\end{align}
Thus, the divergent, surviving contribution emerges from the real part of the response function. To show how this occurs in practice, we now separate the terms that result in current injection. In Eqs.(2)-(3) this amounts to setting $m=n$ in the band summation terms, and taking the clean limit. For legibility, we dropped the $\langle .. \rangle$ brackets identifying expectation values, but they can be easily inferred from context. For Eq.~\eqref{eq:part3} we note that following the appropriate replacement for $\omega_1,\omega_2$, it may be re-written as,
\begin{align}
   -\frac{(2\pi i f_l) O^a_{mn} v^\alpha_{nl} v^\beta_{lm}}{(\varepsilon_l - \varepsilon_n-\hbar\omega_1+i\Gamma)(-\varepsilon_l + \varepsilon_m+\hbar\omega_2+i\Gamma)}   = \frac{(2\pi i f_l) O^a_{mn} v^\alpha_{nl} v^\beta_{lm}}{(\hbar\omega_1 -\varepsilon_{ln}-i\Gamma)(\varepsilon_{mn}+\hbar\Omega)} -\frac{(2\pi i f_l) O^a_{mn} v^\alpha_{nl} v^\beta_{lm}}{(\hbar\omega_2 -\varepsilon_{ml})(\varepsilon_{mn}+\hbar\Omega+i\Gamma)},
\end{align}
yielding again a divergent contribution for $n=m$. We've also used shorthand notation for $\varepsilon_n - \varepsilon_m = \varepsilon_{nm}$.
Assembling the pieces, 
\begin{align}
    O^a = \frac{e^3 A^\alpha (\omega) A^\beta(-\omega)}{2 \Omega \hbar^2} \int \textrm{d}k O^a_{mm}v^\alpha_{ml} v^\beta_{lm} (f_m-f_l) \frac{2 i\Gamma}{(\hbar\omega_0+(\varepsilon_m-\varepsilon_l))^2+\Gamma^2}. 
\end{align}
The factor of $1/2$ emerges from the symmetrization with respect to frequencies.  This expression is easily understood as an analytic extension of the delta function, and we write it, in the $\Gamma \to 0$ limit as,
\begin{align}
i \Omega O^a = -\frac{\pi e^3}{\hbar^2} A^\alpha (\omega_0) A^\beta(-\omega_0)\int \textrm{d}k O^a_{mm}v^\alpha_{ml} v^\beta_{lm} (f_m-f_l) \delta(\hbar\omega_0 + (\varepsilon_m - \varepsilon_l)). 
\end{align}
At this point, it is important to recall the overall symmetrization $(\omega_1 \leftrightarrow \omega_2)$. This symmetrization takes the form $\omega_0 \to -\omega_0$. As such, we introduce this object and perform an exchange of indices with respect to $m,l$. We  get,
\begin{align}
i \Omega O^a = -\frac{\pi e^3}{\hbar^2} \int \textrm{d}k (f_m-f_l)\left[O^a_{mm} v^\alpha_{ml}v^\beta_{lm} A^\alpha (\omega_0) A^\beta(-\omega_0) - O^a_{ll} v^\alpha_{lm}v^\beta_{ml} A^\alpha (-\omega_0) A^\beta(\omega_0)\right]\delta(\hbar\omega_0 + (\varepsilon_m - \varepsilon_l)). 
\end{align}
The current which is divergent in the limit of $\Omega \to 0$ is purely real (the imaginary component, by contrast, vanishes as $\Omega \to 0$). Given that $O^a_{mm}$ is a real quantity (as $O$ is a Hermitian operator), this leaves two possibilities to consider,
\begin{align}
    \textrm{Re}(A^\alpha (\omega) A^\beta(-\omega))\textrm{Re}(v^\alpha_{ml} v^\beta_{lm}), \\
    -\textrm{Im}(A^\alpha (\omega) A^\beta(-\omega))\textrm{Im}(v^\alpha_{ml} v^\beta_{lm}).
\end{align}
The former denotes linear polarized light, while the latter is present only for circularly polarized light. 
Let me now decompose it therefore, as follows,
\begin{align}
    i \Omega O^a_{\textrm{lin}} = -\frac{ \pi e^3}{\hbar^2} A^\alpha (\omega) A^\beta(-\omega)\int \textrm{d}k (f_m-f_l) (O^a_{mm}-O^a_{ll})\left(v^\alpha_{ml} v^\beta_{lm}+v^\beta_{ml} v^\alpha_{lm}\right) \delta(\hbar\omega_0 + (\varepsilon_m - \varepsilon_l))
    \label{eq:withresonance}
\end{align}

We now insert the property of BdG bands into the expressions. Adopting a more convenient notation, the symbols $\varepsilon_{l,m}$ will refer to the unfolded, normal state energies. The BdG states (which include the superconducting gap) are $\xi_h(\mathbf{k}), \xi_e(\mathbf{k})$, i.e., quasielectron and quasihole bands, with their momentum dependence. Certain identities connect the dispersion $\xi_h, \xi_e$ and the observables, or specifically the quasiparticle velocities ($v^\alpha$). For the BdG spectrum, $v^\alpha_{he} = i(\xi_h-\xi_e)\mathcal{A}^\alpha_{he}$ \cite{Tanaka2023}, where $\mathcal{A}^\alpha$ is the particle-hole Berry connection as defined in the main text. The resonance condition in Eq.~\ref{eq:withresonance} implies that $(E_h-E_e) = \omega_0$ meaning that one obtains for linear polarized light (i.e., the case where $\textrm{Re}(A^\alpha A^\beta)$ is non-vanishing),
\begin{align}
    i \Omega O^a_{\textrm{lin}} = -\frac{ \pi e^3}{\hbar^2} E^\alpha (\omega_0) A^\beta(-\omega_0)\int \textrm{d}k (f_h-f_e) (O^a_{h}-O^a_{e})\left(\mathcal{A}^\alpha_{he}\mathcal{A}^\beta_{eh} + \textrm{c.c.} \right) \delta(\hbar\omega + (\xi_e-\xi_h))
    \label{eq:withresonance2}
\end{align}
For concreteness we introduced the occupation factors of $f_h, f_e$ for varying the chemical potential. We also identified $i\omega_0 A^\alpha (\omega_0) = E^\alpha(\omega_0)$.

We can now define $\left(\mathcal{A}^\alpha_{he}\mathcal{A}^\beta_{eh} + \textrm{c.c.} \right)$ as twice band resolved quantum metric $g^{\alpha \beta}_{he} = \frac{1}{2}\mathcal{A}^\alpha_{he}\mathcal{A}^\beta_{eh} + \textrm{c.c.} $ of the BdG bands. The symmetrized and particle-hole conjugated expression appears in the main text.
For circularly polarized light,
\begin{align}
    i \Omega O^a_{\textrm{circ}} = \frac{ \pi e^3}{\hbar^2} E^\alpha (\omega) E^{\beta, *}(-\omega)\int \textrm{d}k (f_h-f_e) (O^a_{h}-O^a_{e})\left(\mathcal{A}^\alpha_{he}\mathcal{A}^\beta_{eh} - \textrm{c.c.} \right) \delta(\hbar\omega + (\xi_e-\xi_h))
    \label{eq:withresonance3}
\end{align}
We designate $\Omega^{\alpha \beta}_{eh} = \mathcal{A}^\alpha_{he}\mathcal{A}^\beta_{eh} - \textrm{c.c.}$ as the Berry curvature of BdG. The sensitivity of the above term to TRS in the BdG ground state depends on the observable, as we detail below.

\subsection{Expressions for the photoconductivity}

Eqs.~\eqref{eq:withresonance}-~\eqref{eq:withresonance2} can be further reduced to the form which appear in the main text. First, we observe that the product $\mathcal{A}^\alpha \mathcal{A}^\beta = Q^{\alpha \beta}$ where $Q^{\alpha \beta}$ is the quantum geometric tensor of the bands. Since $Q^{\alpha \beta} = g^{ab} -\frac{i}{2} \Omega^{ab}$ \cite{Ahn_2021}, we thus arrive at equation (4) of the main text, having reintroduced a factor of $\frac{1}{(2 \pi)^2}$ for the dimensional normalization of the momentum integral, and multiplying this term by 2, in accordance with the prefactor of Eq.~\eqref{eq:withresonance2}. Similarly, we observe that Eq.~\eqref{eq:withresonance3} matches precisely its respective component in $Q^{ab}$. Thus, we recover Eqs. (5)-(6) of the main text.

We now explicitly show how the diagonal component of $O^a = v^a$ is exploited. For $O^a = v^a$, the Hellmann-Feynmann theorem as a function of the perturbation $\mathbf{A}$ in the minimal coupling defined in the main text. Thus, if $\xi_{h/e} = \sqrt{\varepsilon_{e/h}^2(\mathbf{k})+\Delta_{\mathbf{k}}^2}$ and $\xi_{h/e}$ is normal energy of the underlying electron/hole sectors, under minimal coupling with an adiabatic parameter $\mathbf{\lambda}$~\cite{Watanbe2022nonreciprocal}, we have, $\xi_{h/e} = \sqrt{\varepsilon_{e/h}^2(\mathbf{k}+\lambda)+\Delta_{\mathbf{k}}^2}$. The diagonal part of the velocity vertex is then the variation of this quantity with respect to $A$ in the limit $\lambda \to 0$. As a result, 
\begin{align}
    \mathbf{v}_{h/e} = \epsilon_{h/e}\frac{\mathbf{\partial_k}\epsilon_{h/e}}{\xi_{h/e}},
\end{align}
where we used the chain rule $\frac{\partial}{\partial \lambda} = \frac{\partial}{\partial k}\frac{\partial k}{\partial \lambda}$ and $\frac{\partial k}{\partial \lambda} = 1$.
However, $\epsilon_{h/e} = \pm\sqrt{\xi_{h/e}^2-\Delta_{\mathbf{k}}^2}$. Thus, designating $\partial_k \varepsilon_{h/e} = \mathbf{v}_{h/e}$ as the normal state velocity, we directly obtain the definition of the velocity renormalization factor $\mathcal{D}_{h/e} = \sqrt{\xi^2_{h/e}-\Delta^2_\mathbf{k}}\frac{\mathbf{v}}{\xi_{h/e}} = \sqrt{1-\Delta^2_\mathbf{k}/\xi_{h/e}^2} \mathbf{v}^N$. As all injection terms require operator component differences, such as those in Eqs.~\eqref{eq:withresonance2}-\eqref{eq:withresonance3}. As a result, we introduce the symbol, $\mathcal{D}^a_{he} = \sqrt{1-\Delta^2_\mathbf{k}/\xi_{h}^2} v^a_h -\sqrt{1-\Delta^2_\mathbf{k}/\xi_{e}^2} v^a_e$.
% \jp{JP: You have bold vectors on the RHS but not on the LHS....}
The notation $e,h$ indicates quasiparticle and quasihole bands, but for which there exist \textit{additional} degrees of freedom in $H_0$ (as in those leading to inversion symmetry breaking). We stress that in the case of a one-band model, where essentially $v_h = v_e$ all currents, for all operators, vanish, consistent with the standard theory of clean superconductors \cite{Mattis_Bardeen}.

Restoring the normalization of momentum integrals, we recall that they carry a normalization of $(2\pi)^{-d}$ that was thus far ignored. We also note that the definition of the physical velocity operator must include a factor of $\hbar^{-1}$, that is, $v = \frac{1}{\hbar}\partial_k H_0$.

Now, for a linearly polarized injection current we write,
\begin{align}
    \frac{\partial j^a}{\partial t} = -\frac{e^3}{2\pi\hbar}\int \textrm{d}^2k (v^a_h-v^a_e)g^{\alpha \beta}_{he}\delta(\hbar\omega-(\xi_e-\xi_h)),
\end{align}
While for circularly polarized light,
\begin{align}
    \frac{\partial j^a}{\partial t} = \frac{e^3}{4\pi\hbar}\int \textrm{d}^2k (v^a_h-v^a_e)\Omega^{\alpha \beta}_{he}\delta(\hbar\omega-(\xi_e-\xi_h)).
\end{align}
If we use the identity of $Q^{\alpha \beta} = g^{\alpha \beta}-i\Omega^{\alpha \beta}/2$, we can rewrite the total current as,
\begin{align}
    \frac{\partial j^a}{\partial t} = -\frac{e^3}{2\pi\hbar}\int \textrm{d}^2k (v^a_h-v^a_e)Q^{\alpha \beta}_{he}\delta(\hbar\omega-(\xi_e-\xi_h)),
\end{align}
In agreement with Eq.~\eqref{eq:jqtensor}. 

For the out-of-plane polarization, we substitute $p_z$ for $O^a$. For convenience and easy applicability in a bilayer case, we note that for any state, the expectation value of $\langle p_z\rangle$ can be written as the projection on the top layer, denoted as $\mathcal{P}_+$. Its complement is simply $1 - \mathcal{P}_{+}$. This allows us to rewrite the expectation value of any BdG states $e,h$ of the layer polarization as $(d/2)\langle h/e|\mathcal{P}_{+}|h/e\rangle$, where $d$ is the layer separation. With the difference between quasihole and quasiparticle labelled as $\chi^z$, for circularly polarized light we find,
\begin{align}
        \frac{\partial P_z}{\partial t} = \frac{e^3}{4\pi \hbar} \int \textrm{d}^2k (\langle e|\mathcal{P}_{+}|e\rangle-\langle h|\mathcal{P}_{+}|h\rangle)\Omega^{\alpha \beta}_{he} \delta(\hbar \omega -(\xi_e-\xi _h)).
\end{align}
Specifically with $(d)/2(\langle e|\mathcal{P}_{+}|e\rangle-\langle h|\mathcal{P}_{+}|h\rangle) = \chi^z$. Now, this expression can be generalized to an arbitrary number of layers. The total polarization in a multilayer system must be the sum of induced polarization of every layer (assuming states on layers admit a localized description, see above). In this case, we define a projector $\mathcal{P}_l$ as a operator selecting the $l$-th layer from the BdG state. The layer polarization operator is then $\chi^z_{l} = (d)/2(\langle e|\mathcal{P}_{l}|e\rangle-\langle h|\mathcal{P}_{l}|h\rangle) = \chi^z_{l,he}$. We also introduce the summation over $l$ in the final expression, which is,
\begin{align}
        \frac{\partial P_z}{\partial t} = \frac{e^3}{4\pi \hbar} \sum_{l}\int \textrm{d}^2k (\langle e|\mathcal{P}_{l}|e\rangle-\langle h|\mathcal{P}_{l}|h\rangle)\Omega^{\alpha \beta}_{he} \delta(\hbar \omega -(\xi_e-\xi _h)).
\end{align}
Note that reversing the handedness of light, flips the direction of the induced polarization. 
\subsection{Out of normal incidence response}
Thus far, the expressions above focused on terms involved in normal-incident response. Now, we show how the off-normal response is obtained. The electric field is polarized on the $y-z$ plane with the polarization vector $\mathbf{E} = (0,E_y,E_z) = E_0(0,\cos(\phi),\sin(\phi))$. Focusing on the chiral response, that $j^x$, $j^x = \chi^{x;yy} E_y^2 + \chi^{x;yz}E_y E_z^{*}+\chi^{x;yz}E_z E_y^{*}+  \chi^{x;zz} E_z^2$. In a 2D system, the dispersion along $z$ is confined, so we use the dipole operator to represent coupling by the charge dipole to the electric field (see Sub. Sec. ``Polarization operator"). 
%\jp{JP: Broken link}
We define the self-energy,
\begin{align}
    \Sigma(\omega_1) = \mathbf{A}_y(\omega_1)v_y+\mathbf{E}_z(\omega_1)p_z = \Sigma_y(\omega_1) + \Sigma_z(\omega_2),
\end{align}
and look for terms the two contributions that mix them, as the Green's function to second order reads,
\begin{align}
    \notag G(E,\bff{k}) &=  \frac{1}{2}G_r^E \Sigma_z(\omega_1) G_r^{E+\omega_1}\Sigma_y(\omega_2) G_0^{E+\omega_1+\omega_2}+G_0^E \Sigma_z(\omega_1) G_a^{E+\omega_1}\Sigma_y(\omega_2) G_a^{E+\omega_1+\omega_2} + \\ &  G_r^E \Sigma_z(\omega_1) G_0^{E+\omega_1}\Sigma_y(\omega_2) G_a^{E+\omega_1+\omega_2} + (\omega_1,z \Leftrightarrow \omega_2,y).
\end{align}
The result is equations identical to those in Eqs.~\eqref{eq:part1}-\eqref{eq:part3}, with the substitution of $v^\beta$ by $p^z$. For $v^\alpha$, we still have $v^\alpha_{he} = i(\xi_h-\xi_e)\mathcal{A}^\alpha_{he} = i\omega_0 \mathcal{A}^\alpha_{he}$ for states satisfying the resonance condition given above. It follows that $i\omega_0 A^\alpha = E^\alpha$ giving us the equation in the main text. We note that for the polarization as defined above $E_y E_z^{*} = \frac{E_0^2}{2} \sin(2\phi)$ giving the angular dependence on the tilt away from $z$ axis. For $\phi = 0$, the response is purely normal-incident and the chiral current vanishes.

We comment on contribution of terms to the conductivity beyond those that are proportional to $\chi^{x;yz}$. Note that for the mirror-less group $C_4$, the terms $\chi^{x;yy}$ and $\chi^{x;zz}$ vanish identically (see Sec.~\ref{sec:sym}). This leads to the 4-fold pattern upon rotation of the polarization plane in Fig.~\ref{fig:fig4}(a). If the symmetry is reduced to below $C_2$ this can be directly detected by the chiral current, as in Fig.~\ref{fig:fig4}(b). An example of a scenario like this is the emergence of nematicity in the ground state, which naturally breaks rotational symmetry. The chiral response plays an important role, e.g., in the nonlinear transport in the $D_6$ group of twisted bilayer graphene (TBG) \cite{Penaranda2024}.

The reduction in symmetry can also be verified via linear response, as was recently proposed to measure two-component order parameters in Ref.~\cite{sanchezmartinez2025opticalprobestwocomponentpairing}.
% \begin{figure}
%     \centering
%     \includegraphics[width=0.4\linewidth]{figures/j_x_d4.pdf}
%     \includegraphics[width=0.4\linewidth]{figures/j_x_c1.pdf}
%     \caption{Rotational symmetry measured by rotation of the $E_y-E_z$ plane. Left: $C_4$ symmetry of the chiral current. Right: $C_1$ symmetry detected in the chrial current. All fields are normalized to $E_0$.}
%     \label{fig:rotation}
% \end{figure}

\begin{figure}
    \centering
    \includegraphics[width=0.5\linewidth]{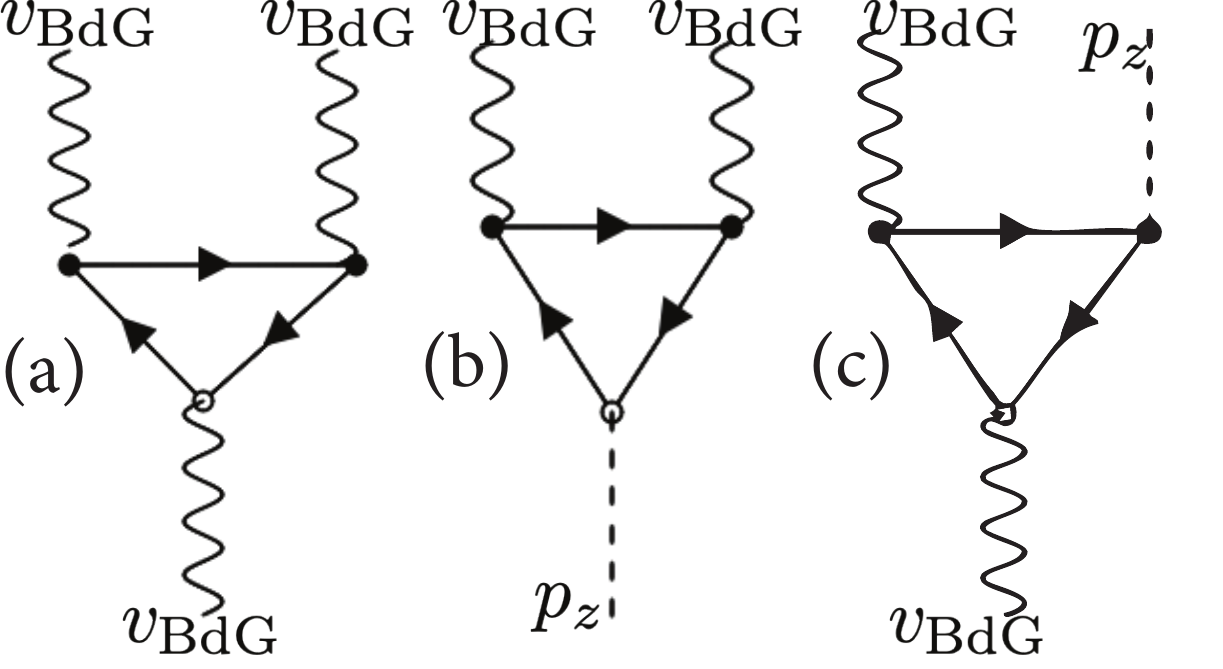}
    \caption{Diagrams contributing to all injection currents. (a) Diagram for normal-incident response, including $v_\textrm{BdG}$ vertices. (b) Diagram for out-of-plane polarization generation. (c) Diagram for the chiral response. For all cases, a filled black dot indicates an incoming photon. A white dot indicates an outgoing response. Incoming vertices are at frequencies $\omega_1,\omega_2$, and symmetrization with respect to these is implied. The real part of diagrams corresponds to linear polarization. Imaginary part is circular polarization.}
    \label{fig:diagrams}
\end{figure}

In all, the contribution of all vertices can be summed up in the following triangle diagrams, which are the \textbf{only} contributions to the injection current. This is plotted in Fig.~\ref{fig:diagrams}. In Fig.~\ref{fig:diagrams}(a) we plot present the triangle diagram of three $v_\textrm{BdG}$ vertices ($v^a$) which corresponds to the in-plane, normal incident current, also presented in Eq.~\eqref{eq:jqtensor}. Fig.~\ref{fig:diagrams}(b) is exclusively generating the polarization injection, Eq.~\eqref{eq:pz}. Fig.~\ref{fig:diagrams}(c) is the chiral current response.

\section{Photoresponse in twisted cuprate superconductors}
\subsection{Minimal model}
The photoresponse of a system of twisted bilayer cuprate superconductors is modeled based on the Hamiltonian introduced in Refs.~\cite{Volkov2023magic,Volkov2023current} for twisted nodal superconductors. Here we consider the following effective BdG Hamiltonian formed at small twist angles,
\begin{align}
    H =\sum_{\mathbf{k}}\Psi_\mathbf{k}^\dagger \left[H(\mathbf{k})\right]\Psi_{\mathbf{k}}.
\end{align}
Here, $\Psi_{\mathbf{k}}$ is the Balian-Werthammer spinor, $\Psi_{k}^\dagger = \left[c^\dagger_{\mathbf{k},l,\uparrow},c_{-\mathbf{k},l,\downarrow},c^\dagger_{\mathbf{k},l,\downarrow},-c_{-\mathbf{k},l,\uparrow} \right]^T$, for layers $l=1,2$ acting in Gor'kov-Nambu ($\tau_i)$ and spin space $(s_i)$. For cuprate superconductors, we assume the order parameter to be for singlet pairing and therefore choose to treat the spin degree of freedom as redundant~\cite{Can_2021}.
$H(\mathbf{k})$ is defined using the notation of Ref.~\cite{Volkov2023current},
\begin{align}
    H(\mathbf{k}) = v_F k_{\parallel} \tau_3 \sigma_0 + v_\perp k_\perp \tau_1\sigma_0-\frac{1}{2}v_\perp Q_N\tau_1  \sigma_3 + t \tau_3 \sigma_1,
    \label{eq:ham_def}
\end{align}
where $\mathbf{Q}_N = (\hat{z} \times \mathbf{K}_N)\theta$ denotes the momentum transfer acquired by rotating the relative positions of the nodal points by an angle $\theta$. The directions $\parallel,\perp$ denote orientations parallel and perpendicular to the nodal momentum $\mathbf{K}_N$, as defined in Ref.~\cite{Volkov2023magic}. As $\mathbf{K}_N$ is taken along the parallel direction, note that $\mathbf{Q}_N$ is strictly along the perpendicular direction in the small twist angle limit. 
%The notation $\mathbf{k}^\theta$ indicates the rotation of the momentum $\mathbf{k}$ by an angle $\theta$. 
If the two layers are stacked with no twist angle ($\theta = 0$), the tunneling between the layers displaces the nodes along the parallel direction. As the twist angle begins to increase ($\theta > 0$), the nodes move towards each other along the parallel direction until they merge at the magic angle, $\theta_{\textrm{MA}}$. The value of $\theta_{\textrm{MA}}$ can be shown explicitly for when the renormalized Fermi velocities (i.e. $v_F$ and $v_\Delta$) vanish at $\theta_{\textrm{MA}} = \frac{2 t }{v_\Delta K_N}$.
Minimal coupling (see main text) in this model would result in a bare vertex, 
\begin{align}
    v^{||} = \tau_3 \frac{\partial H(\mathbf{k})}{\partial k_{||}} = v_F \tau_0 \sigma_0.
    \label{eq:model}
\end{align}
Such a vertex has no off-diagonal components in the BdG basis and cannot lead to optical absorption, consistent with Mattis-Bardeen theory \cite{Mattis_Bardeen}. 
However, Volkov \textit{et al.} \cite{Volkov2023magic} identified contributions to the Hamiltonian arising from the broken circular symmetry of the underlying normal dispersion and gap function $\Delta$. In this case, the leading order corrections takes the form,
\begin{align}
H(\mathbf{k}) \to H(\mathbf{k}) +\delta H_\textrm{non circ.}, 
\quad
\delta H_{\textrm{non circ.}} (\mathbf{k}) = (v_F' \theta k_\perp/2)\tau_3 \sigma_3 - (v_{\Delta}'\theta k_{\parallel}/2)\tau_1 \sigma_3.
\end{align}
Volkov \textit{et al.} have shown that in general $v' \sim v$.
Such a term immediately leads to the emergence of a vertex,
\begin{align}
    v_\perp = (v'_F \theta/2)\sigma_3,
    \label{eq:v_perp}
\end{align}
which mixes the layer degree of freedom and allows optical absorption. This by itself is sufficient for the a current response to linear polarized light. For clarity, we switched to the local coordinate frame near every node. We then carried out the transforation $k_{\parallel} \to k_x, k_\perp \to k_y$. This is the form of the Hamiltonian in the main text.

\subsection{TRS breaking}
The model in Eq.~\ref{eq:model} maintains TRS (see discussion of symmetries below), manifest in the invariance of the dispersion to the operation $\mathbf{k} \to -\mathbf{k}, Q\to -Q,~ H \to \tau_1 \sigma_3 H^{*} \sigma_3 \tau_1$. Since the Berry curvature of the model vanishes, and all other current contributions are absent without TRS breaking, the photoconductivity the model is still strictly zero.
Following Volkov \textit{et al.}, we model TRS breaking by introducing rotation of the Nambu basis by the transformation,
\begin{align}
    \tau_1 \rightarrow \cos\frac{\varphi}{2} \tau_1 - \sin\frac{\varphi}{2} \tau_2\sigma_3
\end{align}
The case of $\varphi = 0$ recovers the original definition of the Nambu convention $\tau$. In the case of an applied interlayer current, $\varphi$ corresponds to the phase difference between the order parameters in both layers \cite{Volkov2023current}. Equally, the same transformation phenomenologically describes the spontaneous realization of the $d+id$ state, with $\varphi$ denotes the strength of TRS breaking. This rotation is essential to the appearance of a finite Berry curvature in the model as well as TRS breaking needed for the observation of finite current.
\subsection{Additional interlayer tunneling}
In the above, only one current (velocity) vertex in the BdG Hamiltonian has interband components allowing for finite absorption. This precludes the possibility of any response to circularly polarized light (in the 2D plane). There exists however an additional contribution in cuprate systems which would permit absorption \cite{Markiewicz2005}, provided there is a non-vanishing orbital splitting at the $\Gamma$ point in the underlying dispersion of the cuprate. Unless symmetry-forbidden (e.g., at a twist angle of $45^o$), this corresponds to an additional term,
\begin{align}
\delta H_{\textrm{tun.}} = (t' k_{||}Q_N) \sigma_1 \tau_3 + \mathcal{O}(k_{\perp}^2).
\end{align}
We ignore corrections at order $k_\perp^2$ as they are subleading compared to the dominant contribution in Eq.~\eqref{eq:v_perp}. Furthermore, the rotation of $\mathbf{K}^\theta = \mathbf{k}^\theta + \mathbf{Q}_N$ introduces a contribution of the form $\mathbf{k}^\theta \mathbf{Q}_N$ which scales as $\sim \theta$. Near the nodal point, it is negligible compared to the tunneling parameter $t$ which remains the largest scale in the problem (they couple in the same channel $\sigma_1 \tau_3$). 
Thus, the leading vertex arising from this,
\begin{align}
    v_{||} = t'Q_N\sigma_1 \tau_0.
\end{align}
The off-diagonal in layer (orbital) basis allows for finite absorption, consistent with Bardeen-Mattis theory. With two independent in-plane directions, circularly polarized light response is allowed.
\section{Symmetries}
\label{sec:sym}
As noted in the main text, the main effect of symmetry is to constrain the independent spatial components of the response tensor related $j^a = \chi^{a;\alpha \beta} E^\alpha E^\beta$. For linear polarized light, we consider the symmetrized component $\chi^{a;\alpha} + \chi^{a;\beta \alpha}$, while for circularly polarized the relevant quantity is the antisymmetric $\chi^{a;\alpha \beta} - \chi^{a;\beta\alpha}$. A rotational group in 2D contains a proper rotation $R_n$ plus an improper rotation in the form of mirrors. These can be combined, and are abelian. As we are dealing with layered systems, we include a 3D rotation which involves exchanging the layers, e.g., $C_{2x}$ or $M_z$. Using the Neumann rule (see main text), the following are the non-zero components of the response for each rotational group (note, we exclude for now operations which rotate the particle-hole basis):
\begin{center}
\begin{table}[h!]
\begin{tabular}{ |p{3cm}||p{7cm}|  }
 %\hline
 %\multicolumn{1}{|c|}{$\chi^{a;\alpha \beta}$}
 %\\
 \hline
 Symmetry & $\chi^{a;\alpha\beta}$ \\ 
 \hline
$C_1$ & All $\chi^{a;\alpha\beta}$ \\ \hline 
$C_2$ & $\chi^{zzz},\chi^{xxz},\chi^{yyz},\chi^{xyz}$ (and permutations) \\ \hline 
$C_3$ & $\chi^{xxx},\chi^{yyy},\chi^{xyz},\chi^{xyy},\chi^{zzz},\chi^{zyy},\chi^{zxx}$ \\ \hline
$C_4, C_6$ & $\chi^{xxz},\chi^{yyz},\chi^{zzz},\chi^{xyz}$ \\ \hline
\end{tabular}
\end{table}
\end{center}
In addition to this, improper rotations such as mirrors and/or $z$ axis rotations. For $C_{2x}$ the non-vanishing components are: $\chi^{xyz}, \chi^{xxy},\chi^{yyz},\chi^{zzy},\chi^{yyy}$. For a mirror reflection: $M_x$, all tensors with a single component perpendicular to the mirror plane, vanish. The combination of the above symmetries imposes constraints \textit{between} non-vanishing components. For exmaple, for $C_3$, $\chi^{xxx} = -\chi^{yyx}, \chi^{yyy} = -\chi^{xxy}$. If one combines a surviving mirror symmetry, as is common in untwisted graphene based systems \cite{Kaplan2022}, $\chi^{yyy} = 0$ and so a single in-plane component defines the response.

For TRS, we note that the effect of spin flip also rotates the particle-hole basis. As a result, $j_\textrm{lin.}$ has to vanish identically. However, $j_\textrm{circ.}$ may be finite, provided the point group symmetry is low enough.
% \subsection{Rotational symmetry breaking}
% The Hamiltonian defined in Eq.~\eqref{eq:ham_def}
%\end{center}
\end{document}